\documentclass{iaus}
\usepackage{graphicx}
\usepackage{epsfig}

\title[Fe/Ni in Mz~3] 
{Fe/Ni ratio in the Ant Nebula Mz 3}

\author[Zhang \& Liu]   
{Y. Zhang \and X.-W. Liu }

\affiliation{Department of Astronomy, Peking University
\break email: zhangy@bac.pku.edu.cn; liuxw@bac.pku.edu.cn}

\pubyear{2006}
\volume{234}  
\pagerange{111--111}
\date{?? and in revised form ??}
\jname{Planetary Nebulae in our Galaxy and Beyond}
\editors{M. J. Barlow \& R. H. M\'endez, eds.}
\begin{document}

\maketitle

\begin{abstract}
We have analyzed the [Fe~{\sc ii}] and [Ni~{\sc ii}] emission lines in the bipolar planetary
nebula Mz~3. We find that the [Fe~{\sc ii}] and [Ni~{\sc ii}] lines arise exclusively
from the central regions. Fluorescence excitation in the formation process
of these lines is negligible for this low-excitation nebula. From the
[Fe~{\sc ii}]/[Ni~{\sc ii}] ratio, we obtain a higher Fe/Ni abundance ratio with respect
to the solar value. The current result provides further supporting evidence for 
Mz~3 as a symbiotic Mira.
\end{abstract}

\firstsection 
\section{Introduction}

The Ant Nebula, Mz~3 is a young bipolar planetary nebula (PN) that consists 
of a bright core, two spherical bipolar lobes and two outer large filamentary 
nebulosities. In recent years, it has been extensively studied. There is 
evidence suggesting that Mz~3 is not a normal PN but contains a symbiotic 
pair at the center (Schmeja \& Kimeswenger 2001).  Zhang \& Liu (2002) find that 
the dense nebular gas at the center may have a different origin from that in 
the extended lobes, and suggest that Mz~3 consists of a giant companion that 
gives rise to the central dense gas and a white dwarf that provides the 
ionizing photons. In the current study, we provide further supporting 
evidence for this scenario by determining the Fe/Ni abundance ratio in Mz~3. 

\section{Analysis}

Our observations have been described by Zhang \& Liu (2002). The slit was
oriented approximately along the nebular major axis and through
the central star. We divided the nebula into two regions. 
As shown in Fig.~1,  [Fe~{\sc ii}] and [Ni~{\sc ii}] lines have been clearly 
observed from  the central region, but are absent 
in the lobes, indicating a possible difference in chemical composition
between the two regions.

\begin{figure}
\centering
\epsfig{file=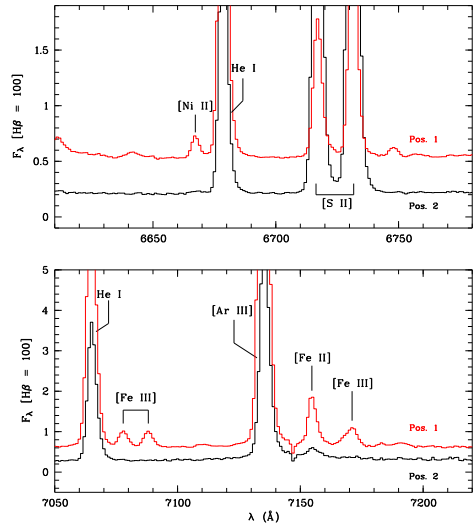, height=7cm,
bbllx=50, bblly=108, bburx=528, bbury=644, clip=, angle=0}
\caption{Spectrum of Mz~3 for a core emission region of nebular radii $r<4$\,arcsec (position 1) and an extended emission region consisting of the two
bright inner lobes on either side of the central star between $4<r<15$\,arcsec
(position 2).
[Fe~{\sc ii}] and [Ni~{\sc ii}] lines arise from position 1 but not from 
position 2. }
\end{figure}

Using the current available atomic data, we have constructed multilevel atomic
 models for Fe$^+$ (159 levels) and Ni$^+$
(76 levels), which were used to obtain the Fe$^+$/H$^+$ and
Ni$^+$/H$^+$ ratios. For our
calculations, continuum fluorescent excitation of [Fe~{\sc ii}] and 
[Ni~{\sc ii}] lines was also considered based on the method of Lucy (1994).
However, we found that fluorescent excitation is not important in this
PN given its low-excitation ($T_\star\sim32,000$\,K) and high density at the center ($\sim10^{6}$\,cm$^{-3}$).
As shown in Table~1, the derived Fe$^+$/H$^+$ and
Ni$^+$/H$^+$ ratios are almost independent of the
dilution factor of continuum radiation, $w$.
Given that Fe and Ni have similar ionization potentials, we
assume ${\rm Fe}/{\rm Ni}={\rm Fe}^+/{\rm Ni}^+$ in deriving the Fe/Ni
abundance ratios.

\begin{table} 
  \begin{center}
  \caption{Fe/Ni ratio for the central emission region of Mz~3 }
  \label{tab:kd}
  \begin{tabular}{lcc}\hline
      $w$  & 0        &   10$^{-3}$  \\\hline
       Fe$^+$/H$^+$   & 9.9$\times10^{-6}$ &  9.2$\times10^{-6}$ \\
       Ni$^+$/H$^+$   &  2.0$\times10^{-7}$ &  1.8$\times10^{-7}$ \\
       Fe/Ni$^{\mathrm{a}}$ & 50 & 51 \\\hline
  \end{tabular}
 \begin{list}{}{}
\item[$^{\mathrm{a}}$] The solar Fe/Ni ratio is 18 (Lodders 2003).
\end{list}
 \end{center}
\end{table}

\section{Discussion}

 Iron serves as the dominant seed of $s$-process which occurs during 
the AGB phase. According to the AGB evolutionary models, iron in the envelop 
is diluted as material is dredged up to the surface from the He innershell 
 (the third dredge-up). Therefore, Fe/Ni ratio is expected to decrease
after the thermally pulsating AGB phase.
Fe and Ni have similar condensation temperatures, thus the amounts of 
depletion onto dust grains, if any, should be comparable. Therefore,
Fe/Ni ratio in a PN should be lower than the solar value.
From the deep spectra presented by Sharpee et al. (2004), we
obtain an Fe/Ni ratio of 4.8 for  PN IC~418, significantly lower than
the solar value, consistent with the predictions of the AGB models.

However, no significant depletion of the Fe/Ni ratio is detected for the 
central dense emission region of Mz~3, as shown in Table~1, suggesting 
that the gas arises 
from winds of  a giant companion where the $s$-process has not occurred
yet. 
Our results show an enhancement of the Fe/Ni ratio relative to the solar value 
and we attribute this to a slightly higher depletion of Ni relative to Fe.

\end{document}